\documentclass[twocolumn,noshowpacs,preprintnumbers,amsmath,amssymb]{revtex4}

\usepackage{verbatim}

\usepackage{graphicx}
\usepackage{dcolumn}
\usepackage{bm}
\usepackage{wrapfig}

\begin{document}

\title{3D characterization of CdSe nanoparticles attached to carbon nanotubes}

\author{Ana B. Hungria}
\email{ana.hungria@uca.es}
\author{Paul A. Midgley}
\affiliation{Department of Materials Science and Metallurgy, University of Cambridge, UK}
\author{Beatriz H. Juarez}
\author{Christian Klinke}
\author{Horst Weller}
\affiliation{Institute of Physical Chemistry, University of Hamburg, Germany}

\begin{abstract} 

The crystallographic structure of CdSe nanoparticles attached to carbon nanotubes has been elucidated by means of high resolution transmission electron microscopy and high angle annular dark field scanning transmission electron microscopy tomography. CdSe rod-like nanoparticles, grown in solution together with carbon nanotubes, undergo a morphological transformation and become attached to the carbon surface. Electron tomography reveals that the nanoparticles are hexagonal-based  with the (001) planes epitaxially matched to the outer graphene layer.

\end{abstract}

\maketitle

The integration of semiconductor nanoparticles and nanoparticle-nanotube composites in solar cells and photovoltaics continues to attract great interest [1-9]. The combined properties of light harvesting nanoparticles together with the high conductivity of carbon nanotubes (CNTs) can lead to improvements in the efficiency of photoelectric devices [10,11]. Well-established synthesis methods developed over the last few years allow the growth of a wide variety of semiconductor nanoparticles with tunable optical properties; examples for efficient light harvesting material in the form of nanoparticles are InP [12], PbS [13,14], CdTe [15], and CdSe [16-18]. The hot injection method [19] has demonstrated its versatility to grow nanoparticles with controlled size and shape. The huge variety of shapes range from nano-dots and nano-rods to nano-stars and nano-tetrapods of one or combined materials. Crystallographic facets of nanoparticles exhibit different reactivity mainly due to the surface morphology and the presence (or absence) of organic ligands capping them [20]. This asymmetry has been used not only to grow anisotropic nanoparticles [21,22] but also to bind metallic or semiconductor nanoparticles [23] or oxides [24] as well as the combination of several semiconductors [18,25,26] on specific sites. Furthermore, the successful doping of nanoparticles has been explained in terms of dopant adhesion to specific facets [27]. This selectivity underlines the importance of elucidating the crystallographic shape of nanoparticles to further understand their reactivity. By means of the hot injection method we demonstrated the possibility to combine nanoparticles with carbon nanotubes. In previous work we described the transformation of rod-like CdSe nanoparticles in a trioctylphosphonic oxide (TOPO)-octadecylphosphonic acid (ODPA) mixture in the presence of both single- and multiwall CNTs [28]. The rod-like nanoparticles obtained at early stages of the reaction undergo a morphological transformation upon interaction with the sp$^{2}$ hybridized carbon lattice and are seen to bond to the nanotube. Here, we present a detailed investigation of the 3D morphology of CdSe nanoparticles and how they bind to carbon nanotubes. We employ high resolution transmission electron microscopy (HREM) and electron tomography to elucidate the structural properties of the composites in three dimensions. Previously, electron tomography has been employed to determine the spatial distribution of nanoparticles inside multiwall CNTs [29] or to study the influence of metallic nanoparticles on the mechanical properties of the CNTs surface [30]. In this work, we used this technique to elucidate the shape of nanoparticles attached to carbon nanotubes. The nanoparticles are dihexagonal pyramids [31] (in this context usually denominated bullet-shaped [13,32,33]) with the (001) planes bonded to the (002) graphitic planes of the nanotubes. The shape evolution was also confirmed by the corresponding optical absorption measurements.

\begin{figure}[!h]
\begin{center}
\includegraphics[width=0.45\textwidth]{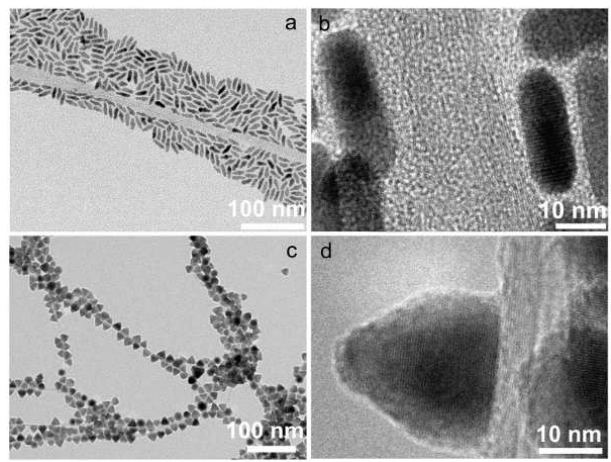}
\caption{\it (a) CdSe rod-like nanoparticles and carbon nanotubes obtained after 1~hour of reaction. (b) A higher magnification image shows a gap in the nanotube-nanoparticle interface corresponding to the length of the ligand molecules. (c) Shape transformed nanoparticles in direct contact to singlewall carbon nanotubes and (d) to a multiwall carbon nanotube.}
\end{center}
\end{figure}

\begin{figure}[!h]
\begin{center}
\includegraphics[width=0.45\textwidth]{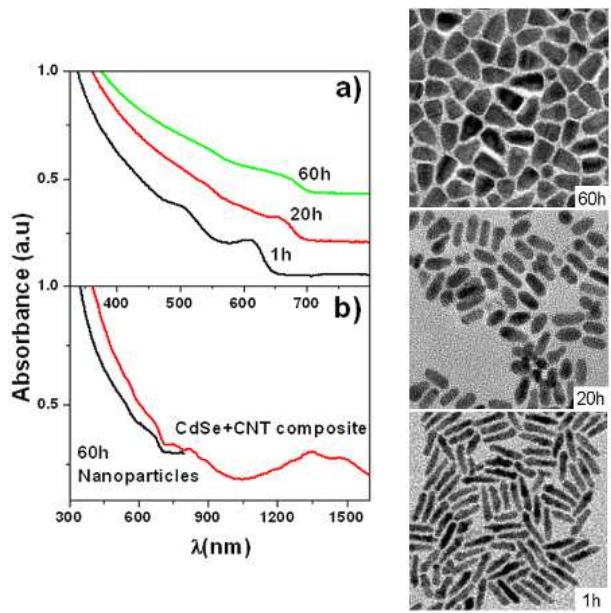}
\caption{\it (a) Evolution of the absorption spectrum of CdSe nanoparticles obtained in the presence of 0.6~mg of singlewall CNTs at different times after 1~h, 20~h, and 60~h. (b) Absorption spectra of CdSe-CNT composites along with that of CdSe nanoparticles after 60~h. Images on the right side show the particles after 1~h, 20~h, and 60~h. Width of the images is 150~nm.}
\end{center}
\end{figure}

The development of the nanoparticles is shown in Figure~1 in which a series of bright field electron micrographs are representative of the different stages of evolution. Figure~1a shows CdSe rod-like nanoparticles together with CNTs after 1~hour of reaction. Nanoparticles arrange around the CNTs due to the drying process of the solvent on the grid. They are separated by about 2~nm, a distance in good agreement with the length of the capping ligands: ODPA and TOP (Figure~1b). It is known that CdSe rod-like nanoparticles grow preferentially parallel to the c-axis with alternating close-packed layers of Cd and Se. Figure~1c and 1d, depict particles obtained after the reaction at 245$^{\circ}$C on singlewall (after 20~h) and multiwall carbon nanotubes (after 60~h), respectively. A shape transformation is seen from rods to pyramids and bonding to the nanotube surface is apparent. We observe a faster transformation and attachment in single than in multiwall CNTs what can be attributed to the stronger graphitic curvature of singlewall type. After the transformation and attachment, the distance of about 2~nm between the nanoparticles and the CNTs can not be observed. As previously reported, the shape transformation is triggered by an etching process of ODPA on Cd sites followed by a consequent release of Se to balance the crystal stoichiometry. In this case, the particles are sitting with the basal plane epitaxially arranged on the graphitic surface of the nanotubes. Whether exclusively van der Waals forces between alkyl chains of the ligands (ODPA, TOPO) are involved in the interaction, cannot be resolved at the present stage of our studies. The broadening of the G-peak in the Raman response addressed in a previous work [28] suggested a possible charge transfer between the nanoparticles and the nanotubes. Thus, an electrostatic interaction between the (probably poor or non-capped) (001) facet and the carbon nanotube lattice is plausible. The nanoparticles can be removed out of the nanotubes by treatment in thiols or pyridine (what suggest a Cd (001) terminated facet rather than a Se one). The proposed mechanism of attachment is described in Reference [28]. 

Once the reaction is completed, the particles in contact with CNTs and the particles that remain in solution have similar shapes. Since CdSe-CNT composites precipitate naturally in toluene, they can be easily separated from the nanoparticles by centrifugation. The temporal evolution of the particles kept in suspension (not in contact with the CNTs) and the CdSe-CNTs composites was followed by means of absorption spectroscopy. Figure~2a shows the evolution of the absorption spectrum of the nanoparticles along with the corresponding TEM images for particles obtained after 1~h, 20~h, and 60~h. The spectrum of the nanoparticles red shifts as a result of particle growth. After 1~h, the nanorods have an average length of 21~nm and are 5~nm in width. The first transition shifts from a wavelength of 617~nm for particles obtained after 1~h to 662~nm for particles obtained after 20~h. The nanoparticle dimensions after this time do not vary significantly in length. However the rods are wider, according to a 1D/2D growing regime [17]. After 60~h, the nanoparticles are pyramidal-shaped and the size distribution broadens. The well-defined absorption peaks (especially the first transition) become smoother and closer to the bulk response for larger and/or more polydispersed particles. The size distribution becomes wider with time as a consequence of Ostwald ripening [34] occurring near equilibrium conditions, leading to the growth of larger particles at the expense of smaller ones. The absorption spectrum of nanoparticles after 60~h of reaction is compared with that of the composite CdSe-CNT in Figure~2b. Singlewall CNTs bear clear absorption peaks between 1000 and 1600~nm which are associated with the first dipole active exciton (E11) and in the range between 500 and 1000~nm which are associated with the second dipole active exciton (E22) of semiconducting CNTs [35]. The spectrum of the composite also shows the absorption edge of the semiconducting CdSe nanoparticles at 670~nm, a value in good agreement with the absorption edge of the nanoparticles not in contact to CNTs. In previous work [28], the direct comparison of the nanotubes spectra before and after the nanoparticles attachment showed not significant deviation neither in peak intensities nor in spectral shifts.

\begin{figure}[!h]
\begin{center}
\includegraphics[width=0.45\textwidth]{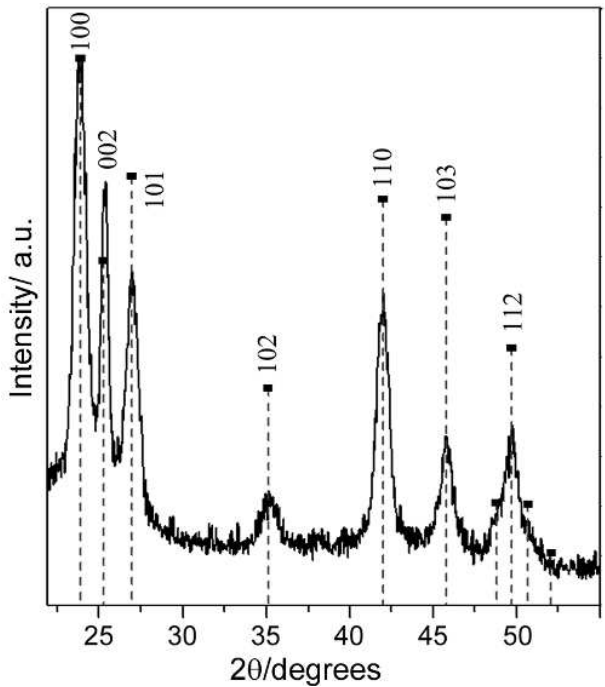}
\caption{\it XRD pattern of CdSe nanoparticles attached to singlewall CNTs. Dashed lines indicate calculated pattern intensities for bulk wurtzite CdSe.}
\end{center}
\end{figure}

\begin{figure}[!h]
\begin{center}
\includegraphics[width=0.45\textwidth]{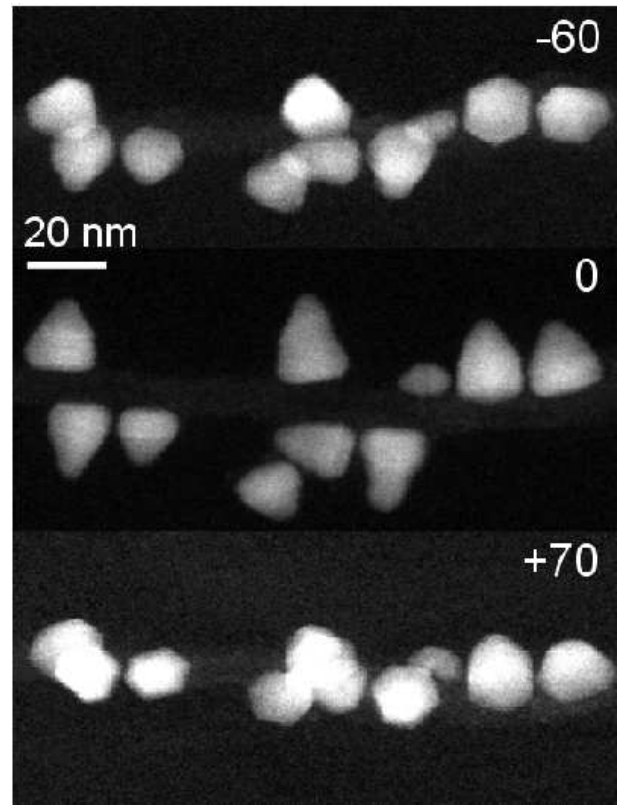}
\caption{\it STEM-HAADF images acquired at different tilt angles showing the hexagonal base of the CdSe particles.}
\end{center}
\end{figure}

The powder X-Ray diffraction (XRD) pattern of the CdSe-CNTs composites is depicted in Figure~3 along with the calculated pattern intensities of bulk CdSe (wurtzite) for comparison [36]. The diffraction pattern shows well-resolved sharp diffraction peaks. Rod-like nanoparticles with the wurtzite structure are distinguished by a sharp (002) peak, which results from the preferential growth of the c-axis of the wurtzite lattice. The crystalline lattice from the transformed, attached, pyramidal-shaped nanoparticles remains hexagonal with a more intense (002) peak (compared to the bulk). The same result is obtained for composites involving different types of CNTs (multiwall or singlewall, regardless of the source). The presence of zinc blende stacking faults along the [001] direction would affect the intensity and width of reflections present in the wurtzite pattern. Broadening of the (103) reflection and the drastic decrease in the (102) reflection have been observed for small nanoparticles ($<$~11~nm)~[37]. A comparison of the intensities of the bulk CdSe pattern and the experimental spectrum depicted in Figure~3, and those shown elsewhere [37], suggest the possibility of this kind of defect being present in the sample. 

High angle annular dark field (HAADF) scanning transmission electron microscopy (STEM) tomography studies were performed to ascertain the three-dimensional geometry of the attached nanocrystals [38,39]. The data for electron tomography were collected by tilting the specimen about a single axis with respect to the electron beam. Figure~4 shows three HAADF-STEM images acquired in an early stage of the tomographic acquisition process at -60$^{\circ}$, 0$^{\circ}$, and +70$^{\circ}$ respectively in which it can clearly be seen that the base of the pyramidal-shaped CdSe nanoparticles attached to the carbon nanotubes is hexagonal. 

\begin{figure}[!h]
\begin{center}
\includegraphics[width=0.45\textwidth]{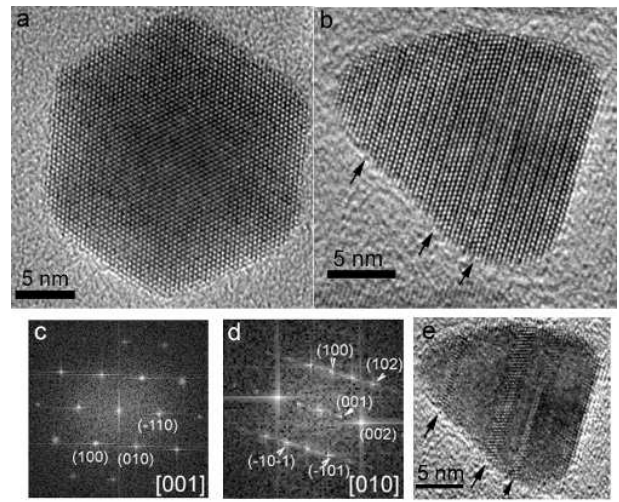}
\caption{\it (a) HRTEM image of a CdSe particle in [001] direction and (b) [010] direction. Note the asymmetry of the pyramidal shape. (c) and (d) show the corresponding power spectra of (a) and (b) respectively. (e) shows a Fourier filtered image of (b) to enhance the zinc blende stacking faults (highlighted with arrows in (b) and (e)).}
\end{center}
\end{figure}

\begin{figure}[!h]
\begin{center}
\includegraphics[width=0.45\textwidth]{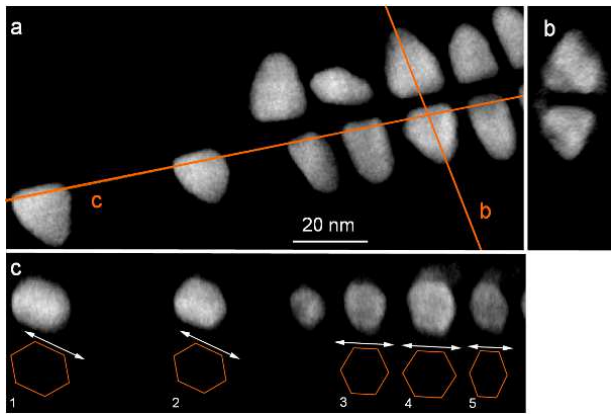}
\caption{\it Mutually perpendicular slices through the reconstructed tomogram. (a) and (b) show the symmetrical distribution of the particles on each side of the nanotube. (c) Depicts the hexagonal shape of the base of the pyramids (highlighted underneath each particle), together with a change in the orientation between particles 1 and 2 and 3 to 5.}
\end{center}
\end{figure}

Figures~5a and 5b show HRTEM images of these hexagonal-based pyramids in plan-view and cross-section respectively, together with the power spectra obtained from these images (5c and 5d correspondingly). Interplanar spacings and angles measured in the FTs correspond to the hexagonal wurtzite structure of CdSe viewed in the [001] (Figure~5a and 5c) and [010] (Figure~5b and 5d) directions. This observation indicates that the (001) planes of the wurtzite structure are in contact with the carbon nanotube, growing parallel to the graphite layers. Nevertheless, although the power spectrum can be initially indexed as corresponding to the wurtzite structure, close inspection of the image depicted in Figure~5d, shows local changes in the structure, indicated by arrows. This kind of defect has been previously described for CdSe [37,40,41], and is formed by the growth of CdSe with a cubic zinc-blende structure. In these zinc-blende layers oriented in the [011] direction, the characteristic zigzag structure of the wurtzite seen in the [010] direction is not present. Fourier filtering of Figure~5b allows the regions of zinc blende structure to be visualized more easily in Figure~5e. 
To complete the characterization of the three-dimensional shape of the nanoparticles and their distribution over the carbon nanotube, a tomographic tilt series of HAADF-STEM images was acquired. Figure~6 shows slices from three perpendicular orientations through the tomographic reconstruction. The carbon nanotube is not visible in the reconstruction due to its low intensity compared to the CdSe nanoparticles. The symmetrical distribution of the CdSe nanoparticles on each side of the carbon nanotube can be seen clearly in both Figure~6a and Figure~6b. The cross sections of most of the particles displayed in Figure~6a show an asymmetry of the pyramidal shape, which was also present in the particle shown in projection in Figure~5b. The shape of the base of the pyramids, highlighted in Figure~6c, is very close to a regular hexagon but slightly distorted in most of the particles. This deformation, consisting of an elongation approximately parallel to the carbon nanotube axis, is not produced by the missing wedge artefact arising from the tomographic reconstruction (related to the limited tilt range), as this would elongate features in the perpendicular direction [42]. Instead, this elongation can be explained as a consequence of the epitaxial ordering of the CdSe nanocrystals with the hexagonal periodicity of the graphite surface of the nanotubes, as shown in the Figure~7. This matching, previously described for CdS crystals grown over graphite [43], could be responsible, at an initial stage, for the stabilization of the planes on the surface of the nanotube. This leads to particles with a wider basal plane than those of a rod-like nanoparticle. The particular shape of the nanotube could favour a higher growth rate ought to coherent interaction of the surfaces. Figure~6c shows the orientation of the particles is slightly tilted respect the tube axis, which could be related with the chiral vector of the outer wall of the nanotube. It can also be noticed a change in the angle of the tilt previously mentioned between particles 1 and 2, on the left hand side of Figure~6c, and particles 3 to 5, on the right. This could be explained by a change in the chirality of the nanotube wall possibly related to the bonding of CdSe nanoparticles to both sides of the nanotube; further studies are needed to confirm this.

A surface-rendered tomographic reconstruction of a CdSe nanoparticle is shown in Figure~8a together with the proposed ideal model for the morphology of these nanoparticles viewed in two different orientations (Figure~8b and 8c). This proposed morphology is a best match with the data obtained from the tomography reconstruction and with the projected images acquired both in TEM and STEM mode. However, a more accurate model could be created if slabs of CdSe zinc-blende structure, which is likely to change the growth rate of individual faces, are included in the simulation.  

\begin{figure}[!h]
\begin{center}
\includegraphics[width=0.45\textwidth]{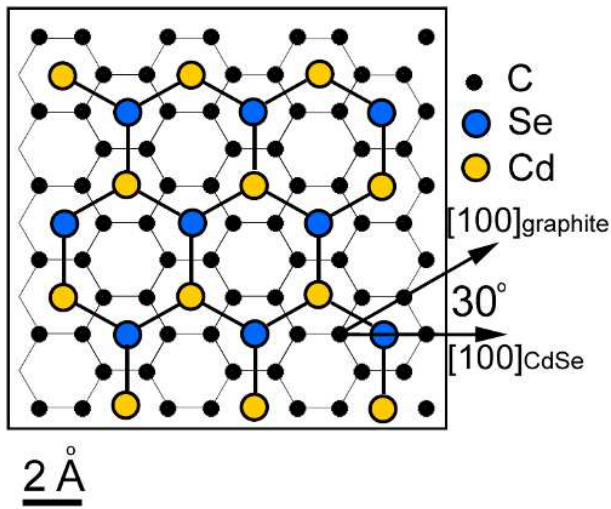}
\caption{\it Simulation of the epitaxial ordering of CdSe over graphite, showing the topmost carbon monolayer (black atoms) and atoms of the first CdSe monolayer, Se (blue) and Cd (yellow). Both graphite and CdSe structures are projected through the [001] direction using Rhodius software~[27].}
\end{center}
\end{figure}

\begin{figure}[!h]
\begin{center}
\includegraphics[width=0.45\textwidth]{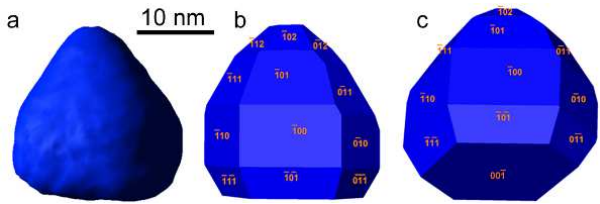}
\caption{\it (a) Surface rendered reconstructed nanoparticle. (b), (c) Model particle at different orientations.}
\end{center}
\end{figure}

In conclusion, the crystallographic structure of pyramidal-shaped CdSe nanoparticles bonded to carbon nanotubes has been elucidated by means of HRTEM and STEM HAADF Electron Tomography. CdSe rod-like nanoparticles grown in solution together with carbon nanotubes undergo a morphological transformation and bond to the carbon surface. Electron tomography investigations reveal that the nanoparticles are hexagonal-based pyramids with the (001) planes epitaxially matched to the carbon lattice. This work points out electron tomography as a powerful tool to elucidate the shape and relative orientation of carbon nanotube-supported nanoparticles, what is crucial to understand the formation of more complex structures.

\subsection*{Experimental Section}

TEM images were acquired on CM20 Philips, JEOL JEM 1011 and CM-300 Philips microscopes. HAADF-STEM tomography was performed on a FEI Tecnai F20 field emission gun transmission electron microscope operated at 200kV. For the tomograms shown in Figures 6 and 8, 64 images were recorded with an acquisition time of 20 seconds every 2$^{\circ}$ from -56$^{\circ}$ to +70$^{\circ}$ using a Fischione ultra-high tilt tomography holder model 2020. The probe size was approximately 0.5~nm in diameter and each HAADF image was recorded with a pixel size of 0.20~nm using a Fischione high angle annular dark field (HAADF) detector. Image acquisition was undertaken using the FEI software package Xplore3D. Images were then aligned sequentially using Inspect 3D. Reconstructions, again with Inspect 3D, were performed using either weighted back-projection (WBP) routines or an iterative routine (SIRT) that constrains the reconstructed volume to match the original images when re-projected back along the original tilt directions. This constraint has the effect of minimising some of the unwanted effects of the limited data sampling and greatly reduces the 'fan' artefact that can be evident in many WBP reconstructions. Voxel projections were constructed in Inspect 3D and surface rendering undertaken using Amira software.

\subsection*{Acknowledgments}

The authors thank Dr. J.A. Pérez-Omil from the University of Cadiz for the use of Rhodius software [44] and Andreas Kornowski for useful discussions. A.B.H. and B.H.J. thank the European Community for their Marie Curie Research Fellowships. P.A.M. thanks the Isaac Newton Trust for funding. The tomography experiments were supported by the IP3 project ESTEEM (Contract number 026019).

\subsection*{References}

{\it [1] Kamat, P.V. Meeting the clean energy demand: Nanostructure architectures for solar energy conversion. Journal of Physical Chemistry C 2007, 111(7): 2834-2860.

[2] Sun, B.Q., Marx, E., and Greenham, N.C. Photovoltaic devices using blends of branched CdSe nanoparticles and conjugated polymers. Nano Letters 2003, 3(7), 961-963.

[3] Mcdonald, S.A., Konstantatos, G., Zhang, S.G., Cyr, P.W., Klem, E.J.D., Levina, L., and Sargent, E.H. Solution-processed PbS quantum dot infrared photodetectors and photovoltaics. Nature Materials 2005, 4(2), 138-U114.

[4] Martinson, A.B.F., Elam, J.W., Hupp, J.T., and Pellin, M.J. ZnO nanotube based dye-sensitized solar cells ZnO nanotube based dye-sensitized solar cells. Nano Letters 2007, 7(8), 2183-2187.

[5] Wang, P., Abrusci, A., Wong, H.M.P., Svensson, M., Andersson, M.R., and Greenham, N.C. Photoinduced charge transfer and efficient solar energy conversion in a blend of a red polyfluorene copolymer with CdSe nanoparticles. Nano Letters 2006, 6(8), 1789-1793.

[6] Plass, R., Pelet, S., Krueger, J., Gratzel, M., and Bach, U. Quantum dot sensitization of organic-inorganic hybrid solar cells. Journal of Physical Chemistry B 2002, 106(31), 7578-7580.

[7] Nozik, A.J. Quantum dot solar cells. Physica E-Low-Dimensional Systems and Nanostructures 2002, 14(1-2), 115-120.

[8] Lee, H., Yoon, S.W., Kim, E.J., and Park, J. In-situ growth of copper sulfide nanocrystals on multiwalled carbon nanotubes and their application as novel solar cell and amperometric glucose sensor materials. Nano Letters 2007, 7(3), 778-784.

[9] Huynh, W.U., Dittmer, J.J., and Alivisatos, A.P. Hybrid nanorod-polymer solar cells. Science 2002, 295(5564), 2425-2427.

[10] Robel, I., Bunker, B.A., and Kamat, P.V. Single-walled carbon nanotube-CdS nanocomposites as light-harvesting assemblies: Photoinduced charge-transfer interactions. Advanced Materials 2005, 17(20), 2458-+.

[11] Sheeney-Haj-Khia, L., Basnar, B., and Willner, I. Efficient generation of photocurrents by using CdS/Carbon nanotube assemblies on electrodes. Angewandte Chemie-International Edition 2005, 44(1), 78-83.

[12] Micic, O.I., Sprague, J.R., Curtis, C.J., Jones, K.M., Machol, J.L., Nozik, A.J., Giessen, H., Fluegel, B., Mohs, G., and Peyghambarian, N. Synthesis and Characterization of Inp, Gap, and Gainp2 Quantum Dots. Journal of Physical Chemistry 1995, 99(19), 7754-7759.

[13] Joo, J., Na, H.B., Yu, T., Yu, J.H., Kim, Y.W., Wu, F.X., Zhang, J.Z., and Hyeon, T. Generalized and facile synthesis of semiconducting metal sulfide nanocrystals. Journal of the American Chemical Society 2003, 125(36), 11100-11105.

[14] Hines, M.A. and Scholes, G.D. Colloidal PbS nanocrystals with size-tunable near-infrared emission: Observation of post-synthesis self-narrowing of the particle size distribution. Advanced Materials 2003, 15(21), 1844-1849.

[15] Yu, W.W., Wang, Y.A., and Peng, X.G. Formation and stability of size-, shape-, and structure-controlled CdTe nanocrystals: Ligand effects on monomers and nanocrystals. Chemistry of Materials 2003, 15(22), 4300-4308.

[16] Peng, Z.A. and Peng, X.G. Formation of high-quality CdTe, CdSe, and CdS nanocrystals using CdO as precursor. Journal of the American Chemical Society 2001, 123(1), 183-184.

[17] Peng, Z.A. and Peng, X.G. Mechanisms of the shape evolution of CdSe nanocrystals. Journal of the American Chemical Society 2001, 123(7), 1389-1395.

[18] Talapin, D.V., Koeppe, R., Gotzinger, S., Kornowski, A., Lupton, J.M., Rogach, A.L., Benson, O., Feldmann, J., and Weller, H. Highly emissive colloidal CdSe/CdS heterostructures of mixed dimensionality. Nano Letters 2003, 3(12), 1677-1681.

[19] Park, J., Joo, J., Kwon, S.G., Jang, Y., and Hyeon, T. Synthesis of monodisperse spherical nanocrystals. Angewandte Chemie-International Edition 2007, 46(25), 4630-4660.

[20] Manna, L., Wang, L.W., Cingolani, R., and Alivisatos, A.P. First-principles modeling of unpassivated and surfactant-passivated bulk facets of wurtzite CdSe: A model system for studying the anisotropic growth of CdSe nanocrystals. Journal of Physical Chemistry B 2005, 109(13), 6183-6192.

[21] Peng, X.G., Manna, L., Yang, W.D., Wickham, J., Scher, E., Kadavanich, A., and Alivisatos, A.P. Shape control of CdSe nanocrystals. Nature 2000, 404(6773), 59-61.

[22] Peng, Z.A. and Peng, X.G. Nearly monodisperse and shape-controlled CdSe nanocrystals via alternative routes: Nucleation and growth. Journal of the American Chemical Society 2002, 124(13), 3343-3353.

[23] Mokari, T., Rothenberg, E., Popov, I., Costi, R., and Banin, U. Selective growth of metal tips onto semiconductor quantum rods and tetrapods. Science 2004, 304(5678), 1787-1790.

[24] Casavola, M., Grillo, V., Carlino, E., Giannini, C., Gozzo, F., Pinel, E.F., Garcia, M.A., Manna, L., Cingolani, R., and Cozzoli, P.D. Topologically controlled growth of magnetic-metal-functionalized semiconductor oxide nanorods. Nano Letters 2007, 7(5), 1386-1395.

[25] Milliron, D.J., Hughes, S.M., Cui, Y., Manna, L., Li, J.B., Wang, L.W., and Alivisatos, A.P. Colloidal nanocrystal heterostructures with linear and branched topology. Nature 2004, 430(6996), 190-195.

[26] Talapin, D. V.; Nelson, J.H.; Shevchenko, E. V.; Aloni, S.; Sadtler, B.; Alivisatos, A. P. Seeded Growth of Highly Luminescent CdSe/CdS Nanoheterostructures with Rod and Tetrapod Morphologies. Nano Letters 2007, 7(10), 2951 - 2959.

[27] Erwin, S.C., Zu, L.J., Haftel, M.I., Efros, A.L., Kennedy, T.A., and Norris, D.J. Doping semiconductor nanocrystals. Nature 2005, 436(7047), 91-94.

[28] Juarez, B.H., Klinke, C., Kornowski, A., and Weller, H. Quantum dot attachment and morphology control by carbon nanotubes. Nano Letters 2007, 7(12), 3564-3568.

[29] Ersen, O., Werckmann, J., Houlle, M., Ledoux, M.J., and Pham-Huu, C. 3D electron microscopy study of metal particles inside multiwalled carbon nanotubes. Nano Letters 2007, 7(7), 1898-1907.

[30] Cha, J.J., Weyland, M., Briere, J.F., Daykov, I.P., Arias, T.A., and Muller, D.A. Three-dimensional imaging of carbon nanotubes deformed by metal islands. Nano Letters 2007, 7(12), 3770-3773.

[31] Kleber, W., Bautsch, H.J., Bohm, J. Einführung in die Kristallographie1998.

[32] Jun, Y.W., Casula, M.F., Sim, J.H., Kim, S.Y., Cheon, J., and Alivisatos, A.P. Surfactant-assisted elimination of a high energy facet as a means of controlling the shapes of TiO2 nanocrystals. Journal of the American Chemical Society 2003, 125(51), 15981-15985.

[33] Nair, P.S., Fritz, K.P., and Scholes, G.D. Evolutionary shape control during colloidal quantum-dot growth. Small 2007, 3(3), 481-487.

[34] Ostwald, W.Z. Phys. Chem. 1900, 34, 495.

[35] Perebeinos, V., Tersoff, J., and Avouris, P. Scaling of excitons in carbon nanotubes. Physical Review Letters 2004, 92(25), 257402 -257405.

[36] Stevenson, A.W., Barnea Z. Acta Crystallographica, B40, 530.

[37] Murray, C.B., Norris, D.J., and Bawendi, M.G., (1993). Synthesis and Characterization of Nearly Monodisperse Cde (E = S, Se, Te) Semiconductor Nanocrystallites. Journal of the American Chemical Society 1980, 115(19), 8706-8715.

[38] Midgley, P.A. and Weyland, M. 3D electron microscopy in the physical sciences: the development of Z-contrast and EFTEM tomography. Ultramicroscopy 2003, 96(3-4), 413-431.

[39] Hernandez, J.C., Hungria, A.B., Perez-Omil, J.A., Trasobares, S., Bernal, S., Midgley, P.A., Alavi, A., and Calvino, J.J. Structural surface investigations of cerium-zirconium mixed oxide nanocrystals with enhanced reducibility. Journal of Physical Chemistry C 2007, 111(26), 9001-9004.

[40] Manna, L.; Scher. E.C.; Alivisatos. A. P. Synthesis of Soluble and Processable Rod-, Arrow-, Teardrop-, and Tetrapod-Shaped CdSe Nanocrystals. Journal of the American Chemical Society 2000, 122, 12700-12706.

[41] Qu, L.H.; Peng, X.G. Control of photoluminescence properties of CdSe nanocrystals in growth. Journal of the American Chemical Society 2002, 124(9), 2049-2055.

[42] Radermacher M, H.W. Proceedings of the 7th European Congr. Electron Microscopy. Den Haag. 1980,

[43] Anderson, M.A., Gorer, S., and Penner, R.M. A hybrid electrochemical/chemical synthesis of supported, luminescent cadmium sulfide nanocrystals. Journal of Physical Chemistry B 1997, 101(31), 5895-5899.

[44] Bernal, S., Botana, F.J., Calvino, J.J., Lopez-Cartes, C., Perez-Omil, J.A., and Rodriguez Izquierdo, J.M. The interpretation of HREM images of supported metal catalysts using image simulation: profile view images. Ultramicroscopy 1998, 72(3-4), 135-164.
}

\clearpage

\end{document}